\documentclass{aastex631}

\usepackage{amsmath}
\usepackage{amsthm}
\usepackage{booktabs}

\theoremstyle{plain}
\newtheorem{theorem}{Theorem}
\newtheorem{lemma}{Lemma}

\newcommand{\G}{\mathcal{G}}
\newcommand{\rb}{\bar{\rho}}
\newcommand{\Sig}{\Sigma}
\newcommand{\be}{\begin{equation}}
\newcommand{\ee}{\end{equation}}

\shorttitle{Gauss--Bonnet zero surfaces in neutron and quark stars}
\shortauthors{Hansraj}

\begin{document}

\title{Where Can Gauss--Bonnet Scalarization Begin in Neutron and Quark Stars?}

\author[0000-0002-8305-7015]{Sudan Hansraj}
\affiliation{Astrophysics Research Centre, School of Mathematics, Statistics and Computer Science,
University of KwaZulu-Natal, Private Bag X54001, Durban 4000, South Africa}
\email{hansrajs@ukzn.ac.za}

\begin{abstract}
In scalar--Gauss--Bonnet gravity, with a coupling $\zeta(\phi)\G$, the Gauss--Bonnet invariant
$\G$ sources the scalar $\phi$ when $\zeta'(0)\neq0$ and sets the sign of the effective mass
squared of its perturbations when $\zeta$ is quadratic, so the surface on which $\G$ changes sign
inside a star organises both the scalar flux and the regions where spontaneous scalarization
can begin.  We locate that surface in general-relativistic stars in $D\geq4$ dimensions.  A
regular static star with continuous density contains a surface $\Sig$ on which $\G$ vanishes
whenever its surface density is below half its mean density, and $\Sig$ is unique when the
density profile steepens outward and $p/\rho$ decreases.  Neutron-star models without density
discontinuities therefore have one, which for the APR4 and SLy equations of state lies just
above nuclear saturation density, about 1.5~km below the surface of a $1.4\,M_{\odot}$ star.  Self-bound stars need not: in the MIT bag model $\G$ is negative
throughout the interior of every stable quark star, as it is for
self-bound matter of any constant sound speed with $0.16<c_{s}^{2}\leq1$.  For a quadratic coupling
the region where scalar perturbations have negative effective mass squared is then, in a
quark star, either the whole interior or the whole exterior, according to the sign of
$\zeta''(0)$, whereas in a neutron star it includes part of the interior for either sign.  In the Newtonian limit $\Sig$ lies where $\rho=\rb/2$ in every dimension, and
for a coupling with $\zeta'(0)\neq0$ the first-order scalar flux through concentric spheres is
largest in magnitude on $\Sig$.
\end{abstract}


\section{Introduction}

The Gauss--Bonnet combination $\G=R_{abcd}R^{abcd}-4R_{ab}R^{ab}+R^{2}$ is, in four dimensions,
the Euler density, but it is also an ordinary local curvature scalar, and its local value
matters wherever it enters the action nonlinearly, as in $f(\G)$ gravity \citep{Nojiri2005}, or
multiplied by a function of a scalar field, as in scalar--Gauss--Bonnet gravity
\citep{DonevaYazadjiev,DonevaYazadjievNS,Silva}.  There a scalar field $\phi$ enters through a
coupling function $\zeta(\phi)$: besides the Einstein--Hilbert term the Lagrangian contains
$\tfrac12\nabla_{a}\phi\nabla^{a}\phi+\zeta(\phi)\G$, in the signature $(+,-,\ldots,-)$ used
here.  Writing $\Box=g^{ab}\nabla_{a}\nabla_{b}$ and $\zeta'=d\zeta/d\phi$, the scalar obeys
\be
\Box\phi=\zeta'(\phi)\,\G .
\label{eq:box}
\ee
This is the scalar equation of \citet{Silva}, where the right-hand side has the opposite
sign because the signature is $(-,+,+,+)$.  When $\zeta'(0)\neq0$, as for a linear coupling
$\zeta=\alpha\phi$ or a dilatonic one, $\G$ sources the scalar; for a weak coupling
$\zeta=\epsilon f(\phi)$, general-relativistic stars are the backgrounds on which the scalar is
computed at first order in $\epsilon$.  When $\zeta(0)=\zeta'(0)=0\neq\zeta''(0)$, as for a
quadratic coupling, general-relativistic stars with $\phi=0$ are exact solutions, and scalar
perturbations about them have effective mass squared $-\zeta''(0)\,\G$.  The sign of $\G$ then
locates the regions in which a tachyonic instability, and with it spontaneous scalarization, can
develop; whether an unstable mode exists is decided by a perturbation analysis of the whole star
\citep{Silva,DonevaYazadjievNS}.  In both cases the sign structure of $\G$ inside
general-relativistic stars is the one that matters.  For a realistic neutron star $\G$ is
negative through most of the interior and turns positive near the surface \citep{Silva,Yamak}, so
that quadratic couplings of opposite sign act in different parts of the star: for one sign the
scalar of a scalarized neutron star is sourced where $\G<0$, for the other where $\G>0$
\citep{Yamak}.  Integrated over a static slice the invariant balances between the star and its
exterior, which is the statement behind the absence of scalar charge for stars in
shift-symmetric scalar--Gauss--Bonnet gravity \citep{YagiSteinYunes}.

That the invariant can change sign inside a star is therefore known.  What has been missing is
a precise account of when the sign change occurs, whether it happens on a single surface, where
that surface lies, and how the local sign structure is tied to the integrated curvature.  This
paper supplies one.  We prove that a regular star with continuous density contains a zero
surface $\Sig$ whenever its surface density is below half its mean density, and that the zero is
unique under two monotonicity conditions satisfied by the standard stellar models; under the
same conditions a surface density above half the mean density leaves no zero, and the sign
changes by a jump at the surface (Theorem~\ref{thm:zero}).  Every star whose density is
continuous and negligible at the surface, as in neutron-star models without density
discontinuities, therefore has a zero surface, whereas no stable quark star of the MIT bag model
has one, and no stable self-bound star of constant sound speed unless the matter is very soft.
For the APR4 and SLy equations of state the zero surface of a $1.4\,M_{\odot}$ star lies about
1.5~km below its surface, although the crust violates the monotonicity conditions.  In the
Newtonian limit $\Sig$ lies where the density equals half the mean density, in every dimension,
which locates it explicitly for polytropes and for the Tolman VII model \citep{Tolman}.  When it
is unique, $\Sig$ is also where the lapse-weighted integral of $\G$ over the enclosed ball is
largest in magnitude, and when $\zeta'(0)\neq0$ that integral is proportional, at first order,
to the scalar flux.  Over a whole slice it vanishes in four dimensions but not above, so that
stars in more than four dimensions carry a scalar charge at first order \citep{Monopoles}.
Geometries in which $\G$ vanishes on whole open regions, and the response of nonlinear $f(\G)$
gravity to a zero of $\G$, are treated in a companion paper \citep{Companion}.

Section~\ref{sec:invariant} gives the invariant inside a star, Sections~\ref{sec:zero}
and~\ref{sec:newton} the zero surface and its Newtonian location, Section~\ref{sec:integral} the
lapse-weighted integral and the scalar flux, and Section~\ref{sec:stars} the relativistic stars,
among them neutron stars with realistic equations of state and self-bound stars.

\section{The invariant inside a star}
\label{sec:invariant}

We use units $G=c=1$, the signature $(+,-,\ldots,-)$ and the field equations
$G_{ab}=8\pi T_{ab}$ in $D$ dimensions, and write a static spherically symmetric metric as
\be
ds^{2}=y(r)^{2}dt^{2}-\frac{dr^{2}}{Z(r)}-r^{2}d\Omega_{D-2}^{2},\qquad Z=1-u ,
\label{eq:metric}
\ee
with $y$ normalised to $1$ at infinity.  For a perfect fluid of energy density $\rho$ and
pressure $p$ the field equations give
\be
u=\frac{16\pi\,r^{2}\rb}{(D-2)(D-1)},\qquad
\rb(r)=\frac{D-1}{r^{D-1}}\int_{0}^{r}s^{D-2}\rho\,ds ,
\label{eq:u}
\ee
so that $\rb$ is the mean density inside radius $r$; in four dimensions $\rb=3m/4\pi r^{3}$ and
$u=2m/r$.  Outside a star of radius $R$ the metric is Schwarzschild--Tangherlini
\citep{Tangherlini}, $Z=y^{2}=1-\mu/r^{D-3}$.  A direct computation of the curvature, with the
field equations, gives
\be
\begin{gathered}
\G_{D}=c_{D}\Big[(\rb-\rho)^{2}-\rho\,(\rho+k_{D}\,p)\Big],\\
c_{D}=\frac{256\pi^{2}(D-3)}{D-1},\qquad k_{D}=\frac{2(D-1)}{D-2}.
\end{gathered}
\label{eq:RW}
\ee
The first term is the Weyl
invariant, $C_{abcd}C^{abcd}=c_{D}(\rb-\rho)^{2}$, and the second comes from the Ricci tensor.
In four dimensions $c_{4}=256\pi^{2}/3$, $k_{4}=3$, and Equation~\eqref{eq:RW} is equivalent to the
expression $\G=48m^{2}/r^{6}-128\pi\rho\,(m+2\pi r^{3}p)/r^{3}$ of \citet{Silva}; the Weyl invariant in terms of the mean density appears in \citet{Garibay},
and, for an isotropic fluid, $\rb-\rho$ is the complexity factor of \citet{Herrera} divided by $4\pi$.  The mean-density form makes the structure plain.  The tidal density
$\rb-\rho=-\tfrac{r}{D-1}\,d\rb/dr$ measures the inhomogeneity of the star.  The matter term is
the product of $\rho$, which sources the curvature of the spatial sections, and $\rho+k_{D}p$;
in four dimensions the latter is the active, or Tolman--Whittaker, density $\rho+3p$ that
sources the lapse.  Hence
\be
\begin{aligned}
\G_{D}(0)&=-c_{D}\,\rho_{c}(\rho_{c}+k_{D}p_{c}),\\
\G_{D}(R^{-})&=c_{D}\,\rb_{R}(\rb_{R}-2\rho_{R}),\\
\G_{D}\big|_{\rm vacuum}&=c_{D}\,\rb^{2},
\end{aligned}
\label{eq:signs}
\ee
where $\rho_{R}=\rho(R^{-})$ and $\rb_{R}=\rb(R)$ are surface values and the surface value of
$\G_{D}$ uses $p(R)=0$.  At a regular centre with $\rho_{c}>0$ and $\rho_{c}+k_{D}p_{c}>0$ the
matter term wins and $\G_{D}<0$; outside the star there is nothing but Weyl curvature and
$\G_{D}>0$.

\section{Existence and uniqueness of the zero surface}
\label{sec:zero}

\begin{figure}[t]
\centering
\includegraphics[width=0.62\textwidth]{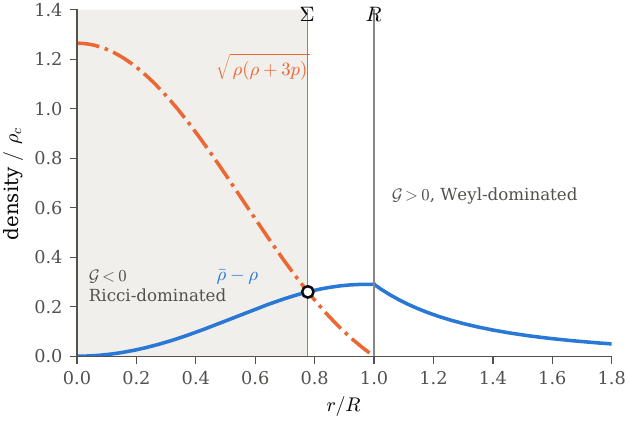}
\caption{The locator \eqref{eq:locator} in an $n=1$ polytrope with $p_{c}/\rho_{c}=0.2$
($2M/R=0.401$).  The zero surface lies where the tidal density $\rb-\rho$, which continues
outside as $3M/4\pi r^{3}$, crosses $\sqrt{\rho(\rho+3p)}$; here $r_{0}/R=0.7771$.}
\label{fig:balance}
\end{figure}

For a nonincreasing density $\rb\geq\rho$, so for $p\geq0$ Equation~\eqref{eq:RW} factorises with a
positive second factor and $\G_{D}$ has the sign of
\be
F(r)=\frac{\rb}{\rho}-1-\sqrt{1+\frac{k_{D}\,p}{\rho}} .
\label{eq:F}
\ee
Indeed $\G_{D}=c_{D}\,\rho^{2}F\big(F+2\sqrt{1+k_{D}p/\rho}\,\big)$, and the second factor,
$\rb/\rho-1+\sqrt{1+k_{D}p/\rho}$, is at least $1$.  A zero surface is therefore located by
\be
\rb=\rho+\sqrt{\rho\,(\rho+k_{D}\,p)}\,,
\label{eq:locator}
\ee
the radius at which the tidal density has grown to the geometric mean of the two material
factors (Figure~\ref{fig:balance}).

\begin{lemma}
\label{lem:ratio}
If $s(r)=d\ln\rho/d\ln r$ is nonincreasing, then $\rb/\rho$ is nondecreasing.
\end{lemma}

\begin{proof}
For $t<r$, monotonicity of $s$ gives $\ln\rho(r)-\ln\rho(t)\geq s(r)\ln(r/t)$, so
$\rho(t)\leq\rho(r)(t/r)^{s(r)}$.  If $s(r)>-(D-1)$, inserting this in Equation~\eqref{eq:u} gives
$\rb(r)\leq(D-1)\rho(r)/(D-1+s(r))$, that is, $d\ln\rb/d\ln r=(D-1)(\rho/\rb-1)\geq s(r)$; if
$s(r)\leq-(D-1)$ the same inequality holds because $\rho/\rb\geq0$.  Hence
$d\ln(\rb/\rho)/d\ln r\geq0$.
\end{proof}

\begin{theorem}[Zero surface]
\label{thm:zero}
Let a static perfect-fluid star in $D\geq4$ dimensional general relativity have a regular
centre, $\rho$ and $p$ continuous on $[0,R)$, $\rho>0$ nonincreasing, $p\geq0$ and $p(R)=0$.
\begin{enumerate}
\item[(i)] If $\rho_{R}<\rb_{R}/2$, then $\G_{D}$ vanishes somewhere in $(0,R)$.
\item[(ii)] If moreover $d\ln\rho/d\ln r$ is nonincreasing and $p/\rho$ is strictly decreasing,
the zero $r_{0}$ is unique, with $\G_{D}<0$ for $r<r_{0}$ and $\G_{D}>0$ for $r>r_{0}$.
\item[(iii)] If $\rho_{R}\geq\rb_{R}/2$ under the hypotheses of (ii), then $\G_{D}<0$
throughout $(0,R)$ and there is no interior zero; for $\rho_{R}>\rb_{R}/2$ the sign changes by a
jump at $r=R$.
\end{enumerate}
\end{theorem}

\begin{proof}
(i) follows from $\G_{D}(0)<0<\G_{D}(R^{-})$, by Equation~\eqref{eq:signs} and continuity.  Under
(ii) the first term of Equation~\eqref{eq:F} is nondecreasing by Lemma~\ref{lem:ratio} and the square
root is strictly decreasing, so $F$ is strictly increasing and has at most one zero.  (iii) is
the case $F(R^{-})\leq0$.
\end{proof}

Case (iii) is not a zero surface at the boundary.  When $\rho_{R}>\rb_{R}/2$,
$\G_{D}(R^{-})<0<\G_{D}(R^{+})$ is a discontinuous change of sign, and the intermediate value
theorem supplies no point at which the invariant vanishes; at equality the interior limit is
zero.  Continuity of the density inside the star matters for the same reason.  At an interior
density discontinuity, such as the interface of a first-order phase transition, $\G_{D}$ jumps
and can change sign there without vanishing, so that the zero surface is replaced by a
sign-changing interface.  For a cold neutron star with an iron surface $\rho_{R}/\rb_{R}$ is of
order $10^{-14}$, and (i) applies whenever the density is continuous inside.

\section{The Newtonian zero surface}
\label{sec:newton}

For $p\ll\rho$ the locator \eqref{eq:locator} becomes
\be
\rho=\tfrac12\rb ,
\label{eq:half}
\ee
in every dimension; equivalently, the second invariant of the Newtonian tidal tensor vanishes
there.  For a polytrope $p=K\rho^{1+1/n}$ in $D-1$ spatial dimensions,
$\rho=\rho_{c}\theta^{n}(\xi)$ with $r=a\xi$, where $a$ is the usual length scale and $\theta$
solves the Lane--Emden equation $\xi^{2-D}(\xi^{D-2}\theta')'=-\theta^{n}$ with $\theta(0)=1$ and
$\theta'(0)=0$.  Since this gives $\rb=-(D-1)\rho_{c}\theta'/\xi$, Equation~\eqref{eq:half} reads
$2\xi\theta^{n}+(D-1)\theta'=0$.  Table~\ref{tab:newton} lists its roots.  In four dimensions the
zero tracks the core rather than the radius: the root $\xi_{0}$ of this condition falls only
from $2.56$ to $2.02$ as $n$ runs from $0.5$ to $3$, while the first zero $\xi_{1}$ of $\theta$,
the stellar radius, grows from $2.75$ to $6.90$.  For the Tolman VII density
$\rho=\rho_{c}(1-r^{2}/R^{2})$ \citep{Tolman}, Equation~\eqref{eq:half} gives the closed forms
\be
\begin{gathered}
x_{0}=\frac{r_{0}}{R}=\sqrt{\frac{D+1}{D+3}},\\
\frac{m(r_{0})}{M}=\frac{(D+1)x_{0}^{D-1}-(D-1)x_{0}^{D+1}}{2},
\end{gathered}
\label{eq:tolman}
\ee
that is, $\sqrt{5/7}$ and $0.862$ in four dimensions.  For uniform density
$\rb=\rho$ and there is no zero.

\begin{table*}[tb]
\caption{The Newtonian zero surface $\rho=\rb/2$ of polytropes and of the Tolman VII model, with
the fraction of the mass it encloses.  Polytropes in $D-1$ spatial dimensions have a finite
radius only for $n<(D+1)/(D-3)$.}
\label{tab:newton}
\setlength{\tabcolsep}{9pt}
\begin{tabular}{lcccccc}
\toprule
& \multicolumn{2}{c}{$D=4$} & \multicolumn{2}{c}{$D=5$} & \multicolumn{2}{c}{$D=6$}\\
\cmidrule(lr){2-3}\cmidrule(lr){4-5}\cmidrule(lr){6-7}
$n$ & $r_{0}/R$ & $m(r_{0})/M$ & $r_{0}/R$ & $m(r_{0})/M$ & $r_{0}/R$ & $m(r_{0})/M$\\
\midrule
0.5 & 0.9317 & 0.928 & 0.9412 & 0.918 & 0.9483 & 0.911\\
1   & 0.7832 & 0.809 & 0.7971 & 0.767 & 0.8085 & 0.732\\
1.5 & 0.6387 & 0.714 & 0.6348 & 0.632 & 0.6254 & 0.550\\
2   & 0.5089 & 0.638 & 0.4647 & 0.510 & 0.3844 & 0.359\\
3   & 0.2923 & 0.524 & -- & -- & -- & --\\
\midrule
Tolman VII & 0.8452 & 0.862 & 0.8660 & 0.844 & 0.8819 & 0.830\\
\bottomrule
\end{tabular}
\end{table*}

\section{The lapse-weighted integral and the scalar flux}
\label{sec:integral}

\begin{figure*}[tb]
\centering
\includegraphics[width=\textwidth]{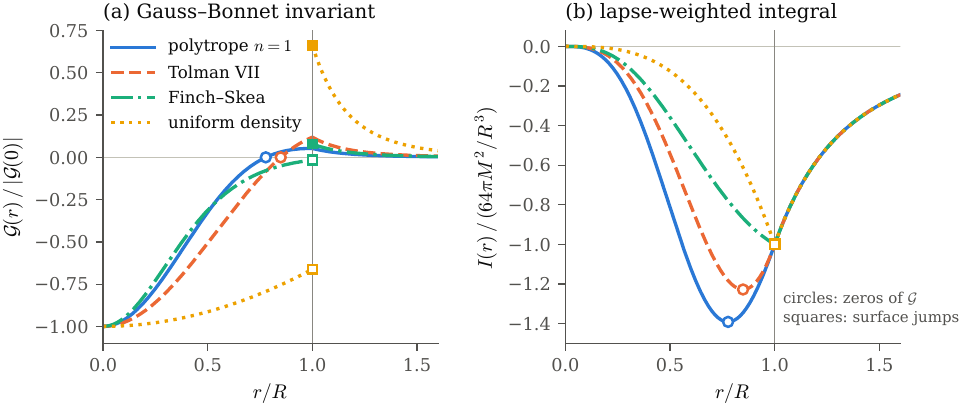}
\caption{(a) The invariant, normalised to the magnitude of its central value, for an $n=1$
polytrope ($2M/R=0.40$), Tolman VII ($0.30$), Finch--Skea ($CR^{2}=1.5$, $2M/R=0.60$) and
uniform density ($0.40$), continued into the exterior.  Circles mark zeros of $\G$.  Finch--Skea
and uniform density have none: their invariant jumps at $r=R$ between the interior limit (open
square) and the exterior value (filled square).  The interior joins the exterior continuously
only when $\rho_{R}=0$.  (b) The lapse-weighted integral \eqref{eq:Idef}, in units of
$64\pi M^{2}/R^{3}$.  Every curve reaches $-1$ at the surface and joins the common exterior
$-(R/r)^{3}$.  Circles mark smooth minima, at the zero surfaces; squares mark the corner minima
at $r=R$ of the two models without a zero.}
\label{fig:gallery}
\end{figure*}

The local signs are tied together by an integral.  Let
\be
I(r)=\int_{B_{r}}\G_{D}\,N\,dV_{\gamma}
=\Omega_{D-2}\int_{0}^{r}\G_{D}\,\frac{y\,s^{D-2}}{\sqrt{Z}}\,ds ,
\label{eq:Idef}
\ee
where $B_{r}$ is the ball bounded by the sphere $S_{r}$ of areal radius $r$ in a static slice
$t=$ const, $N=y$ is the lapse, $\gamma_{ij}$ the Riemannian (positive-definite) metric of the
slice, so that $ds^{2}=N^{2}dt^{2}-\gamma_{ij}dx^{i}dx^{j}$, $dV_{\gamma}$ its volume element and
$\Omega_{D-2}$ the area of the unit $(D-2)$-sphere.  This is the integral of $\G_{D}\sqrt{-g}$ per
unit Killing time,
fixed by the normalisation of the lapse at infinity; it is not the proper-volume integral of
$\G_{D}$, and it is not an Euler characteristic.  By construction
\be
I'(r)=\Omega_{D-2}\,r^{D-2}\,\G_{D}\,\frac{y}{\sqrt{Z}} ,
\label{eq:Iprime}
\ee
so wherever $\G_{D}$ is continuous its zeros are stationary points of $I$, and a change of sign
from negative to positive is a local minimum.  Under Theorem~\ref{thm:zero}(ii) the integral
therefore decreases from $I(0)=0$ to its global minimum on $\Sig$ and increases beyond it.
Under (iii) the minimum sits at $r=R$, where $I$ has a corner because $I'$ jumps with $\G_{D}$;
that minimum is not a zero of the invariant (Figure~\ref{fig:gallery}).

In four dimensions $\G\sqrt{-g}$ is a total derivative, and a direct computation of the
curvature of the metric \eqref{eq:metric} gives it explicitly,
\be
r^{2}\,\G_{4}\,\frac{y}{\sqrt{Z}}=-8\,\big(\sqrt{Z}\,u\,y'\big)' ,
\label{eq:total4}
\ee
as is familiar from the Noether-symmetry treatment of $f(\G)$ gravity \citep{Bahamonde}.
Integrating from the regular centre, where $u\,y'=O(r^{3})$, and using the field equations, one
finds
\be
I(r)=-64\pi\,\frac{y}{\sqrt{Z}}\,\frac{m\,(m+4\pi r^{3}p)}{r^{3}}\qquad(D=4),
\label{eq:I4}
\ee
which is negative at every radius for $m>0$ and $m+4\pi r^{3}p>0$ and vanishes at the centre and
at infinity.  The interior contribution, $I(R)=-64\pi M^{2}/R^{3}$, is therefore cancelled
exactly by the exterior one, the balance found in \citet{YagiSteinYunes}.  Above four
dimensions the cancellation fails, because the Gauss--Bonnet density of the spatial slices no
longer vanishes identically, and the integral over the whole slice is strictly negative for
every star with $\rho+p>0$ \citep{Monopoles}.

\subsection{Scalar flux}
The integral has a direct meaning in scalar--Gauss--Bonnet gravity.
Take a weak coupling $\zeta=\epsilon f(\phi)$ with $f'(0)\neq0$, linear or dilatonic, and
$\phi\to0$ at infinity.  To first order in $\epsilon$ the scalar is then a test field on the star
of general relativity: its stress tensor and the Gauss--Bonnet terms of the metric
equations are of order $\epsilon^{2}$.  (A dilatonic coupling has $f(0)\neq0$, which above four
dimensions switches on an Einstein--Gauss--Bonnet term already at order $\epsilon$; its effect
on $\phi$ is of order $\epsilon^{2}$ and does not enter here.)  The static form of
Equation~\eqref{eq:box} is
\be
D_{i}(N D^{i}\phi)=-\zeta'(0)\,N\G ,
\label{eq:static}
\ee
with $D_{i}$ the covariant derivative of $\gamma_{ij}$, and integrating over $B_{r}$ gives the
flux through $S_{r}$,
\be
\oint_{S_{r}}N D^{i}\phi\,dS_{i}=\Omega_{D-2}\,r^{D-2}\,y\sqrt{Z}\,\phi'(r)=-\zeta'(0)\,I(r),
\label{eq:flux}
\ee
the second form holding for the metric \eqref{eq:metric}.  So $I(r)$ is, up to the coupling, the
scalar flux through the sphere of radius $r$.  When the zero surface is unique, as under
Theorem~\ref{thm:zero}(ii), the flux is therefore largest in magnitude on $\Sig$, and when
$\G_{D}$ has no zero it is largest at the surface: beyond either point $I$ increases towards its
limit at infinity, which is zero in four dimensions and negative above.  In four dimensions
Equation~\eqref{eq:I4} gives $\phi'=16\,\zeta'(0)\,m\,(m+4\pi r^{3}p)/(r^{5}Z)$, which outside the star
falls off as $r^{-5}$: there is no monopole at this order, in agreement with
\citet{YagiSteinYunes}.  All of this is perturbative.  At higher order the metric responds
to the scalar, and for a nonlinear coupling a charge may then appear even in four dimensions:
the first-order cancellation says nothing about it.

\section{Relativistic stars}
\label{sec:stars}

\subsection{Models and method}
The polytropes obey $p=K\rho^{1+1/n}$ with $\rho$ the total energy
density \citep{Tooper}.  In units with $\rho_{c}=1$, and with $\sigma=p_{c}/\rho_{c}=K\rho_{c}^{1/n}$,
one has $\rho=\theta^{n}$ and $p=\sigma\theta^{n+1}$, and the Tolman--Oppenheimer--Volkoff
equations, extended to $D$ dimensions \citep{PonceDeLeon}, are integrated outward from the
regular-centre series with an adaptive Runge--Kutta scheme at relative tolerance $10^{-11}$,
the mean density being integrated directly.  The surface is located as the event $\theta=0$,
and zeros of $\G$ by bracketing on a grid of 4000 points followed by Brent's method.  The values
of $\G$ from Equation~\eqref{eq:RW} agree with those from the metric to $10^{-14}$, and for $n=1$,
$\sigma=0.2$ the quantities $2M/R=0.40144061$, $r_{0}/R=0.77711305$ and $m(r_{0})/M=0.81270798$
are unchanged in all quoted digits when the tolerance is varied between $10^{-8}$ and
$10^{-12}$.  In four dimensions we take $n=0.5$, $1$, $1.5$, $2$ and $3$ and, for each index, 28
values of $\sigma$ spaced logarithmically from $10^{-4}$ to the causal value $n/(n+1)$, at which
the central sound speed reaches that of light: 140 models.  These are equilibrium
configurations, and not all of them are stable.  With $\rho_{c}=1$ the mass at fixed $K$ is
proportional to $\mu\,\sigma^{-n(D-3)/2}$.  In four dimensions it increases up to the causal
value for $n=0.5$, has its first maximum, where radial stability is lost \citep{HadzicLin}, at
$\sigma=0.42$, $0.20$ and $0.097$ for $n=1$, $1.5$ and $2$, and decreases from the Newtonian
limit for $n=3$, so that 101 of the 140 models lie on stable branches.  The first mass maximum
and the compactness maximum are located by Brent's method.  Theorem~\ref{thm:zero} assumes
equilibrium only, and we keep the remaining models to show that the zero surface does not depend
on stability.  The Tolman VII and Finch--Skea models are integrated inward from $p(R)=0$; the
uniform-density star is an analytic control and not part of a causal sequence.  The
integrations and the symbolic and numerical checks of all identities were carried out with
Python (SciPy, SymPy, mpmath) scripts written and run with the assistance of Claude (Anthropic;
Claude Opus 5.5 and Claude Fable 5.2), and the author checked the derivations.  A separate
Mathematica implementation, also written with that assistance, was run by the author and
reproduces, in spherical symmetry, the identities, as well as Table~\ref{tab:newton}, the
figures and the values quoted for the polytropes.  The scripts are provided as ancillary files
\citep{Ancillary}.

\begin{figure*}[tb]
\centering
\includegraphics[width=\textwidth]{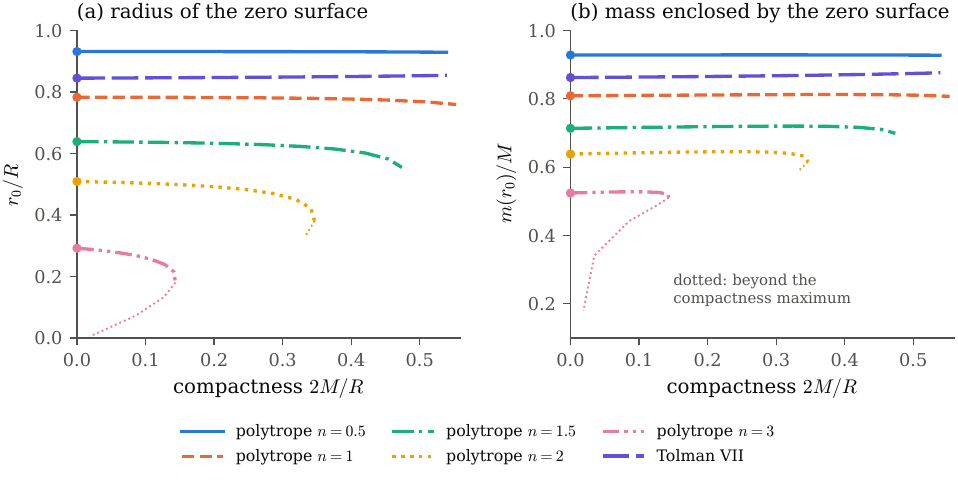}
\caption{Location of the zero surface against compactness for the 140 four-dimensional
polytropes described in the text and for Tolman VII stars up to their causal limit: (a) its
radius, (b) the mass it encloses.  Solid and dashed lines follow each polytropic sequence from
$p_{c}/\rho_{c}=10^{-4}$ to its compactness maximum; dotted lines continue it to the causal value
$n/(n+1)$.  Dots at zero compactness are the Newtonian values of Table~\ref{tab:newton}.}
\label{fig:location}
\end{figure*}

\subsection{Polytropes}
Every one of the 140 four-dimensional polytropes, five indices with 28
central pressures each, has exactly one zero.  The hypotheses of Theorem~\ref{thm:zero}(ii) hold
for all of them except two $n=3$ models beyond the compactness maximum, which nonetheless have a
single zero; the condition on $p/\rho=\sigma\theta$ always holds.  On each branch up to the
compactness maximum the zero moves inward with compactness (Figure~\ref{fig:location}(a)) and the
fraction of the mass it encloses changes little: $m(r_{0})/M$ ranges over $0.928$--$0.929$,
$0.807$--$0.813$, $0.699$--$0.720$, $0.616$--$0.645$ and $0.511$--$0.528$ for $n=0.5$, $1$,
$1.5$, $2$ and $3$.  Beyond the maximum the fraction falls, to $0.588$ for $n=2$ and to $0.179$
for $n=3$ at the causal limit (dotted in Figure~\ref{fig:location}(b)).  On the stable branches
alone the ranges are $0.809$--$0.813$, $0.714$--$0.720$ and $0.638$--$0.645$ for $n=1$, $1.5$
and $2$, and unchanged for $n=0.5$.  This near constancy holds along each sampled sequence below
its compactness maximum, not for the equation of state in general, and it does not mean that
the zero surface follows the same matter from one star to the next.

\subsection{Realistic equations of state}
Polytropes are only a proxy for neutron-star matter,
so we have also integrated the APR4 \citep{APR} and SLy \citep{DouchinHaensel} equations of state
in the piecewise-polytropic form of \citet{Read}, three polytropes in the core
joined to a four-piece fit of the SLy crust.  The structure equations are integrated in the
logarithm of the pressure, which resolves the crust, and two integrators give the same values to
all quoted digits.  The maximum masses, $2.188$ and $2.048\,M_{\odot}$, and the radii of
$1.4\,M_{\odot}$ stars, $11.32$ and $11.71$~km, agree with those of the fits in
\citet{Read} to $0.1\%$.  Neither equation of state satisfies the hypotheses of
Theorem~\ref{thm:zero}(ii): in the crust layer around neutron drip the adiabatic index of the fit
is below $1$, so that $p/\rho$ increases outward, and $d\ln\rho/d\ln r$ increases outward
wherever the index increases outward across a junction of the fit.  Yet each of 30 stars from
$1.0\,M_{\odot}$ to the maximum mass has exactly one zero.  In a $1.4\,M_{\odot}$ star it lies at $r_{0}=9.86$~km for APR4 and $10.15$~km
for SLy, $1.46$ and $1.56$~km below the surface, where the rest-mass density is $1.14$ and
$1.04$ times the nuclear saturation density $2.7\times10^{14}$~g\,cm$^{-3}$, and it encloses
$92.4\%$ and $92.0\%$ of the mass.  In the maximum-mass stars $r_{0}/R=0.897$ and $0.864$ and
$m(r_{0})/M=0.914$ and $0.884$.  These are the regions, $\G<0$ inside $\Sig$ and $\G>0$ outside
it, that source the scalar of scalarized APR4 stars for the two signs of the coupling
\citep{Yamak}.

\subsection{Higher dimensions}
Theorem~\ref{thm:zero} and the Newtonian location
\eqref{eq:half} hold in every dimension.  We have integrated polytropes with $n=1$ and $1.5$ in
five and six dimensions, with 30 values of $\sigma$ spaced logarithmically from $10^{-3}$ to
$n/(n+1)$.  Each has exactly one zero, which tends to its Newtonian value as the compactness
falls; in six dimensions, for example, $r_{0}/R=0.8074$ at $\sigma=0.01$ for $n=1$, against
$0.8085$ in Table~\ref{tab:newton}.  The local picture is therefore the same above four
dimensions; the global one is not, since the integral over the whole slice no longer vanishes
\citep{Monopoles}.

\subsection{Tolman VII}
Here $\rho_{R}=0$, and Theorem~\ref{thm:zero} gives exactly one zero at
every compactness.  Up to the causal limit $2M/R=0.540$ it moves outward only from $0.8452R$ to
$0.8544R$, enclosing between $86\%$ and $88\%$ of the mass.

\subsection{Finch--Skea}
For $\rho=C(3+w)/8\pi(1+w)^{2}$ with $w=Cr^{2}$ \citep{FinchSkea} one
finds
\be
(\rb-\rho)^{2}-\rho^{2}=\frac{3C^{2}(w-3)}{64\pi^{2}(1+w)^{3}} .
\label{eq:FS}
\ee
Since pressure only lowers $\G$ and vanishes at the surface, an interior zero requires
$CR^{2}>3$, that is $2M/R>3/4$.  Causality fails first: the central sound speed reaches that of
light at $CR^{2}=2.611$ ($2M/R=0.723$), and the largest sound speed in the interior, attained near
$r=0.43R$, already does so at $CR^{2}=2.544$ ($2M/R=0.718$).  Every causal Finch--Skea star
therefore has $\G<0$ throughout its interior and no zero surface; as in case (iii) of
Theorem~\ref{thm:zero}, the sign changes by a jump at $r=R$ from a negative interior value to the
positive exterior one.  The uniform-density star behaves in the same way at any compactness.

\subsection{Quark stars}
Self-bound stars show that the absence of a zero surface is not an
artefact of idealised models.  We take the four-dimensional MIT bag model of strange quark
matter with massless, non-interacting quarks, $p=(\rho-4B)/3$ with $B>0$
\citep{Witten,FarhiJaffe,AlcockFarhiOlinto,HaenselZdunikSchaeffer}, whose surface density is
$\rho_{R}=4B$.  The equation of state is homogeneous in $\rho$ and $B$, so in units of $4B$ the
sequence of stars is the same for every $B$.  Along the stable branch, which ends at the first
maximum of the mass \citep{HadzicLin}, $\rho_{R}/\rb_{R}$ falls from $1$ to $0.5639$ at the
maximum, where $\rho_{c}=4.8126\times4B$ and $2M/R=0.5415$; it drops below $\tfrac12$ only on the
unstable branch, at $\rho_{c}=7.13\times4B$.  The surface value alone does not exclude a zero
inside, so we evaluated Equation~\eqref{eq:F} at 6000 radii in each of 80 stable stars with $\rho_{c}$
between $1.0001$ and $4.8126$ times $4B$.  In all of them $\rb/\rho\leq1.774$ throughout, and
since $p\geq0$ makes the square root in Equation~\eqref{eq:F} at least $1$, $F\leq\rb/\rho-2<0$.
Every stable bag-model quark star therefore has $\G<0$ throughout its interior and no zero
surface, and the sign changes by a jump at the surface.  For $B^{1/4}=145$~MeV the maximum mass
is $2.0\,M_{\odot}$, at $R=10.9$~km.

Nor does the result rest on the bag model.  Self-bound matter with constant sound speed,
$p=c_{s}^{2}(\rho-\rho_{s})$, contains it at $c_{s}^{2}=1/3$, $\rho_{s}=4B$, and gives the same
sequence of stars for every $\rho_{s}$.  For 13 values of $c_{s}^{2}$ between $0.16$ and the
causal value $1$ we evaluated Equation~\eqref{eq:F} in the same way in 60 stable stars each.  Along
every stable branch $\rho_{R}/\rb_{R}$ falls monotonically from $1$ to its value at the maximum
mass, which rises from $0.500$ at $c_{s}^{2}=0.16$ through $0.564$ at $c_{s}^{2}=1/3$ to $0.684$
at $c_{s}^{2}=1$; $\rb/\rho$ is largest at the surface and stays below $2$, and no stable star
has a zero surface.  Only for softer matter, $c_{s}^{2}<0.160$, does $\rho_{R}$ fall below
$\rb_{R}/2$ on the stable branch, and then only in stars close to the maximum mass, which do have
a zero surface.

A neutron-star model with continuous density, whose surface density is negligible, always has a
zero surface, by Theorem~\ref{thm:zero}(i).  With a quadratic coupling, for which scalar
perturbations have effective mass squared $-\zeta''(0)\,\G$, the region where it is negative is
therefore, for a quark star, either the whole interior or entirely outside the star, according
to the sign of $\zeta''(0)$, whereas a neutron star contains regions of both kinds; whether an
unstable mode exists requires a separate perturbation calculation.  In scalar--tensor theory, in
which quark-star scalarization has also been studied \citep{MehraeenNazdik}, the trace
$\rho-3p$ plays the role of $\G$; for massless quarks it equals $4B$ throughout the star, and the
corresponding region is the whole interior for one sign of the coupling and empty for the other.

\section{Discussion}
\label{sec:discussion}

The sign structure of the Gauss--Bonnet invariant inside a star is fixed by a single comparison,
between the Weyl term set by the mean density and the product of the two material sources.  It
follows that, for continuous density, a zero surface exists whenever the surface density is below
half the mean density, that it is unique for the usual monotone profiles, that it sits at
$\rho=\rb/2$ in the Newtonian limit in every dimension, and that for stars with a finite surface
density it may be absent altogether, the sign then changing by a jump.  That case is physical:
no stable quark star of the MIT bag model has a zero surface, nor any stable self-bound star of
constant sound speed with $0.16<c_{s}^{2}\leq1$, whereas every neutron-star model with continuous
density has one, which for realistic equations of state lies just above nuclear saturation
density, about 1.5~km below the surface of a $1.4\,M_{\odot}$ star.  In scalar--Gauss--Bonnet
gravity $\Sig$ separates the regions in which a quadratic coupling of either sign gives scalar
perturbations a negative effective mass squared \citep{Silva}.  In four-dimensional nonlinear
$f(\G)$ gravity the correction to Einstein's equations has an algebraic part proportional to
$(f-\G f_{\G})\,g_{ab}$ and a part built from the Hessian of $f_{\G}$ \citep{Nojiri2005}.  On
$\Sig$, where $\G=0$, the first reduces to $f(0)\,g_{ab}$, which vanishes when $f(0)=0$, while
the Hessian, there equal to
$f_{\G\G}(0)\nabla_{a}\nabla_{b}\G+f_{\G\G\G}(0)\nabla_{a}\G\,\nabla_{b}\G$, does not vanish in
general.  A zero of the invariant therefore does not switch the correction off; an explicit
evaluation is given in the companion paper \citep{Companion}.  Above four dimensions the field
equations contain, in addition, the Gauss--Bonnet field tensor multiplied by $f_{\G}$.

When it is unique, the zero surface is also where, for a coupling with $\zeta'(0)\neq0$, the
scalar flux through concentric spheres is largest at first order; far from the star the flux
vanishes in four dimensions but not above, where stars carry a scalar charge \citep{Monopoles}.
Our analysis assumes static, spherically symmetric, isotropic stars in general relativity; the
quark-star results use the bag model and self-bound matter of constant sound speed; and the near
constancy of the enclosed mass fraction is a property of the sampled sequences rather than a
universal one.  Anisotropic stars, whose zero surface is no longer located by Equation~\eqref{eq:RW},
and stars in Einstein--Gauss--Bonnet gravity itself, are natural extensions.

\begin{acknowledgments}
The author thanks Prof.\ Dawood Kothawala (IIT Madras) and the late Prof.\ Naresh Dadhich
(IUCAA) for early discussions on this project.
\end{acknowledgments}

\software{NumPy, SciPy, SymPy, mpmath, Matplotlib, Mathematica}

\section*{Data availability}
The data that support the findings of this article are openly available \citep{Ancillary}.

\section*{AI disclosure statement}
Claude (Anthropic; Claude Opus 5.5 and Claude Fable 5.2) was used to draft and edit parts of the text and to
write the Python and Mathematica scripts that check the calculations; the author checked the derivations
and every number quoted.  Responsibility for the final manuscript text lies entirely with the author.


\end{document}